\documentclass[12pt]{article}
\usepackage{latexsym}
\usepackage{amsmath}
\usepackage{indentfirst}
\usepackage{pstricks} 
\usepackage{cite}
\usepackage{graphicx}

\numberwithin{equation}{section}

\long\def\ca#1\cb{}
\def\outl#1{\par{\medskip\noindent\hspace*{0.1cm}\bf
      \mathversion{bold}#1\mathversion{normal}\smallskip} }
   \def\xa{} \def\xb{}  

 \def\outl#1{}\def\xa{}\def\xb{}

\ca
 \def\outl#1{\par{\medskip\noindent\hspace*{.5cm}\bf
      \mathversion{bold}#1\mathversion{normal}\smallskip} }
 \long\def\xa#1\xb{} 
  
\cb

\newcommand{\ad}{^\dagger }

\newcommand{\becs}{\begin{cases}}
\newcommand{\bem}{\begin{matrix}}

\newcommand{\dya}[1]{|#1\rangle\langle#1|}

\newcommand{\encs}{\end{cases}}
\newcommand{\enm}{\end{matrix}}

\newcommand{\ket}[1]{|#1\rangle }

\newcommand{\lra}{\leftrightarrow }

\newcommand{\mat}[1]{\left(\begin{matrix}#1\end{matrix}\right)}

\newcommand{\mte}[2]{\langle#1|#2|#1\rangle }

\newcommand{\ot}{\otimes }

\newcommand{\ra}{\rightarrow }
\newcommand{\Ra}{\Rightarrow }

\newcommand{\tm}{\times }
\newcommand{\Tr}{{\rm Tr}}

\newcommand{\vbl}{\,\boldsymbol{|}\,}

\newcommand{\CC}{{\mathcal C}}
\newcommand{\DC}{{\mathcal D}}

\newcommand{\HC}{{\mathcal H}}

\newcommand{\XC}{{\mathcal X}}

\newcommand{\dl}{\delta }

\newcommand{\sg}{\sigma }

\begin{document}

\title{Quantum Measurements and Contextuality}
\author{Robert B. Griffiths\thanks{Electronic address: rgrif@cmu.edu}\\
Department of Physics\\
Carnegie Mellon University\\
Pittsburgh, PA 15213}

\date{Version of 5 August 2019}
\maketitle
\xa

\begin{abstract}
  In quantum physics the term `contextual' can be used in more than one way.
  One usage, here called `Bell contextual' since the idea goes back to Bell, is
  that if $A$, $B$ and $C$ are three quantum observables, with $A$ compatible
  (i.e., commuting) with $B$ and also with $C$, whereas $B$ and $C$ are
  incompatible, a measurement of $A$ might yield a different result (indicating
  that quantum mechanics is contextual) depending upon whether $A$ is measured
  along with $B$ (the $\{A,B\}$ context) or with $C$ (the $\{A,C\}$ context).
  An analysis of what projective quantum measurements measure shows that quantum
  theory is Bell noncontextual: the outcome of a particular $A$ measurement
  when $A$ is measured along with $B$ would have been exactly the same if $A$
  had, instead, been measured along with $C$.

  A different definition, here called `globally (non)contextual' refers to
  whether or not there is ('noncontextual') or is not ('contextual') a single
  joint probability distribution that simultaneously assigns probabilities in a
  consistent manner to the outcomes of measurements of a certain collection of
  observables, not all of which are compatible. A simple example shows that
  such a joint probability distribution can exist even in a situation where the
  measurement probabilities cannot refer to properties of a quantum system, and
  hence lack physical significance, even though mathematically well-defined. It
  is noted that the quantum sample space, a projective decomposition of the
  identity, required for interpreting measurements of incompatible properties
  in different runs of an experiment using different types of apparatus has a
  tensor product structure, a fact sometimes overlooked.
\end{abstract}

\xb
\tableofcontents
\xa

\xb
\section{Introduction \label{sct1}}
\xa

\xb
\outl{Two (of many) uses of `contextual': Bell and Abramsky et al. }
\xa

The terms `context' and `contextual' are used with a variety of different
meanings. In this article they refer to the concepts as they appear in quantum
physics, in particular quantum foundations; there is no attempt to relate them
to usages which occur in other fields, such as psychology. Unfortunately, even
in the restricted domain of quantum foundations one finds more than one usage
and, what is equally unfortunate, carelessness on the part of some authors
whose definitions are not clearly stated, and sometimes hard to extract from
the context (to use a different sense of that word) in which their discussions
appear. All of which adds to the general confusion which has given quantum
foundations a bad reputation among physicists working in other disciplines. In
this paper two quite distinct usages of `(non)contextual' will be discussed.
One is due to Bell, and appears in older work on quantum foundations; for this
I will use the term `Bell (non)contextual'. It is discussed in Secs.~\ref{sct2}
to \ref{sct4}. The other is frequently found in more recent work and for it I
use the term `global (non)contextual'. A precise definition, motivated by work
of Abramsky et al.\ \cite{AbBM17}, is presented in Sec.~\ref{sct5}.
There are undoubtedly other definitions to be found in the literature, but 
in the present work I limit myself to these two.

\xb
\outl{To relate `contextual'\& `measurement', must understand `measurement'}
\xa

\xb
\outl{First and second measurement problems discussed in Sec.~\ref{sct3}}
\xa

As the title suggests, my aim is to relate `(non)contextual' to the topic of
quantum measurements. Here there is enormous confusion, as exemplified by the
well-known \emph{measurement problem}. Alas, `measurement' is another term
often used in quantum foundations without a clear meaning, so of necessity any
discussion of how it relates to contextuality must be based on some
understanding of, if not solution to, the measurement problem. In fact there
are (at least) two measurement problems. The first, widely discussed, problem
is how to think about the macroscopic outcome of a measurement, traditionally
referred to as a pointer position, since unitary time evolution often results
in a quantum superposition (Schr\"odinger cat) of possible measurement
outcomes. The second, less discussed but equally important, is how to infer
from the macroscopic outcome the earlier microscopic property that the
apparatus was designed to measure. Both problems are discussed in
Sec.~\ref{sct3}. The second is the one most intimately connected to notions of
quantum contextuality.

\xb
\outl{CH approach employed in this analysis}
\xa

The approach employed in the remainder of this paper is, without apology, based
upon the \emph{consistent histories} (CH), or \emph{decoherent histories}
interpretation of quantum mechanics. It is, so far as I know, the only
interpretation presently available which can provide a complete solution to
both measurement problems. There is no room in a short article to reproduce its
essential features; the reader will find an accessible overview in
\cite{Grff14b}, and a helpful introduction to the way it deals with the
aforementioned measurement problem(s) in \cite{Grff11b,Grff15,Grff17b}. The
conclusion in Sec.~\ref{sct4} is that quantum theory is Bell
\emph{noncontextual}, in agreement with earlier work in \cite{Grff13b} and in
Sec.~V~E of \cite{Grff17b}. The notion of `global contextuality' (my
terminology) motivated by \cite{AbBM17} is discussed in Sec.~\ref{sct5}. There
is an overall summary in Sec.~\ref{sct6}.

\xb
\section{Bell Contextual \label{sct2}}
\xa

\xb
\outl{Bell contextual: $A$ commutes with $B$ and $C$; $BC\neq CB$}
\xa

So far as I know, the term `contextual' as used in quantum foundations goes
back to an early paper by Bell \cite{Bll661}.%
\footnote{Bell did not use the term ``contextual,'' which was only later
  introduced to represent his idea; see Shimony's discussion in \cite{Shmn09b},
  and in particular the remarks included in reference 8 in the bibliography of
  \cite{Shmn09b}.} %
It is not altogether easy to follow Bell's presentation, but I believe the
general idea is the following; see, e.g., the discussion in Sec.~VII of
\cite{Mrmn93}. Let $A$, $B$, and $C$ be three quantum observables---for present
purposes an 'observable' is any Hermitian operator on a (finite-dimensional)
Hilbert space---and assume that $A$ commutes with $B$ and with $C$, but $B$
does not commute with $C$. Hence according to a principle of standard
(textbook) quantum mechanics, it is possible to measure $A$ along with $B$, or
$A$ along with $C$, but one cannot measure all three, $A$, $B$, and $C$ at the
same time, i.e, in a single experimental run. Bell noted that for this reason a
measurement of $A$ along with $B$ requires a different apparatus than a
measurement of $A$ along with $C$, and, with good reason, asked the question:
would the value obtained for $A$ be the same when measured with $B$ as when
measured with $C$? If the answer is `Yes' then quantum mechanics (or whatever
theory one is using to analyze this situation) is \emph{noncontextual}, whereas
if the answer is `No', or at least `No' in certain cases, quantum mechanics (or
the theory in question) is \emph{contextual} in the sense that the measurement
outcome for $A$ depends upon whether it is measured along with $B$, the
$\{A,B\}$ context or together with $C$, the $\{A,C\}$ context.

\xb
\outl{Issues: Measurements=macro outcomes or micro properties? Counterfactual:
$A$ in a particular $AB$ run $\ra$ same value as if it had been an $AC$ run?}
\xa

This distinction seems clear enough until one gives it careful thought and
finds that there are conceptual traps hidden inside what looks at the outset
like a straightforward definition. One of these is found in the term
`measured': what is a quantum measurement? Does one mean a macroscopic
measurement \emph{outcome}, a pointer position if we use the picturesque,
albeit archaic, terminology of quantum foundations? Or is it the prior
microscopic property inferred from the pointer position? Let us assume the
latter: the apparatus pointer indicates that observable $A$ had the value
$A=a_1$ in a particular run of the experiment, one in which $B$ was measured at
the same time as $A$ using what we might call the $AB$ apparatus. If, instead,
\emph{in this particular run}, $A$ had been measured along with $C$ using the
(necessarily different) $AC$ apparatus, would the $A$ pointer have ended up
once again indicating $A=a_1$?

\xb
\outl{Not the same as: marginal prob distr for $A$ the same in $AB$ and $AC$
  runs} 
\xa

Note that this is not at all the same question as asking whether the
\emph{probability distribution} for $A$ outcomes during a set of repeated $AB$
measurements, all starting with the same initial state for the particle, would
be the same as that of the $A$ outcomes during a similar set of $AC$
measurements. Standard textbook quantum mechanics tells us that if we start
with a particular initial state the marginal distribution of $A$ outcomes
computed from the joint distribution of two commuting observables $A$ and $B$
will be the same as that computed from the joint distribution of $A$ and $C$.
The question raised by Bell was not the identity of distributions, but the
identity of outcomes in the \emph{same} run. Since $A$, $B$, and $C$ cannot be
measured simultaneously, this is a \emph{counterfactual} question: $A$ was
measured with $B$; what \emph{would have been} the $A$ outcome \emph{had it}
(contrary to fact) been measured along with $C$?

\xb
\outl{``Qm Bell contextual?'' raises subtle issues in Qm foundations}
\xa

Thus the question of whether quantum mechanics is or is not Bell contextual
has embedded in it some issues in quantum foundations concerning which it is
safe to say there is no general agreement. My position regarding them will
emerge in the next section; all I can ask of the sceptical reader is to pay
attention to the arguments and try and assess their validity with an open mind.

\xb
\section{Quantum Measurements \label{sct3} }
\xa

\xb
\outl{$A=A\ad = \sum_j a_j P_j$; $\{P_j\}$ is a PDI}
\xa

\xb
\outl{Measurement $P-j\ket{\psi}=\ket{\psi} \ra \ket{\Psi}$ with 
$M_j\ket{\Psi} = \ket{\Psi}$;  PDI $\{M_k\}$ $\lra$ pointer positions} 
\xa

Consider a projective measurement on a system, hereafter thought of as a
\emph{particle}, of  an observable $A=A\ad$ with spectral representation
\begin{equation}
 A = \sum_j a_j P_j,\quad \sum_j P_j = I_p
\label{eqn1}
\end{equation}
where the $a_j$ are the distinct eigenvalues of $A$, thus $a_j\neq a_k$ if
$j\neq k$, and the $P_j$ are projectors, $P_j = P_j\ad = P_j^2$, which are
mutually orthogonal: for $j\neq k$, $P_j P_k=0$. The collection $\{P_j\}$ of
orthogonal projectors which sum to the identity $I_p$ on the Hilbert space
$\HC_p$ of the particle is a \emph{projective decomposition of the identity} or
PDI. The measuring process is assumed to be such that if the particle is
initially in some state $\ket{\psi}$ with the (microscopic) property $P_j$,
i.e., $P_j\ket{\psi}=\ket{\psi}$, then its interaction with the measurement
apparatus will, by unitary time evolution, lead to a state $\ket{\Psi}$ which
lies in a subspace, with projector $M_j$, of the apparatus Hilbert space
$\HC_m$, i.e., $M_j\ket{\Psi}=\ket{\Psi}$. The subspace $M_j$ corresponds to
the (macroscopic) apparatus pointer being in the position $j$. Since the
different pointer positions are macroscopically distinct, we can safely assume
that $M_jM_k=0$ for $j\neq k$, and by adding an additional projector $M_0$ to
cover all other possibilities (e.g., the apparatus has broken down), we can
assume that $\{M_j\}$ is a projective decomposition of the identity $I_m$ of
the apparatus Hilbert space. Note that we are thinking of the $\{M_j\}$ as
referring to a later time when the measurement is over and the particle has
disappeared, or else has become a very small part of what we call the
apparatus---this gets around any need to `collapse' the particle wavefunction
at the end of the measurement. However, if one adopts von Neumann's measurement
model in which the particle continues to exist as a separate entity when the
measurement is over, so one is dealing with a Hilbert space $\HC_p\ot\HC_m$,
one need only replace $M_j$ in the preceding discussion with $I_p\ot M_j$, a
projector which signifies that the apparatus pointer is in position $j$, and
provides no information about the state of the particle. (The idea that a
measurement should ``collapse'' the particle wavefunction is only relevant if
one is considering the subsequent history of the particle after the measurement
is over. See the discussion in Sec.~IV~C of \cite{Grff17b}.)

\xb
\outl{1st measurement problem. Macro superposition. $\ket{\Psi}$ is
  pre-prob; $p_j = \mte{\Psi}{M_j}$ }
\xa

Next we need to dispose of the two measurement problems. The first arises when
the initial particle state is some superposition of states corresponding to
different eigenvalues of $A$, say
\begin{equation}
 \ket{\psi} = \sum_j r_j \ket{\phi_j},\quad
 P_k\ket{\psi_j} = \dl_{jk} \ket{\phi_j},
\label{eqn2}
\end{equation}
with at least two of the $r_j$ unequal to zero. Then unitary time evolution
will, as is well-known, lead to a later apparatus state $\ket{\Psi}$ which
no longer falls in just one of the pointer subspaces $\{M_j\}$, so it is
a macroscopic quantum superposition, or in popular terminology a Schr\"odinger
cat. What shall we do with it? Let us follow the advice of Born and use the
Schr\"odinger wave $\ket{\Psi}$ to assign a probability 
\begin{equation}
 p_j = \mte{\Psi}{M_j}
\label{eqn3}
\end{equation}
to each measurement outcome. When used in this manner I refer to $\ket{\Psi}$
as a \emph{pre-probability}, i.e., something used to generate a probability
distribution, as in \eqref{eqn3}, to be carefully distinguished from the
projector $[\Psi] = \dya{\Psi}$ that represents a hard-to-interpret
\emph{quantum property}, the weird `cat state'. 

\xb 
\outl{Sample space. Two PDIs $\{P_j\}$, $\{Q_k\}$ (in)compatible. Single
  framework rule}
\xa

The CH interpretation treats quantum mechanics, following Born, as a stochastic
theory in which time development must be understood using probabilities. But as
in classical physics, probabilities require a \emph{sample space} of mutually
exclusive possibilities, one and only one of which is thought to take place in
a particular run of the experiment. In quantum theory a sample space is always
a PDI. The orthogonality of the projectors ensures that only one, and the fact
that they sum to the identity means at least one, of these possibilities
will occur in any particular situation. In classical physics the choice of a
sample space is usually quite straightforward, and if two or more spaces are of
interest in some situation it causes no difficulty, as one can always combine
them. But in quantum mechanics one needs to make different choices depending on
what aspect of a situation one is interested in. Thus let $\{P_j\}$ and
$\{Q_k\}$ be two PDIs for the same system. If and only if they are
\emph{compatible},
\begin{equation}
 P_jQ_k = Q_kP_j \text{ for all $j$ and $k$}
\label{eqn4}
\end{equation}
can there be a common refinement, a PDI consisting of all the nonzero $P_jQ_k$;
otherwise they are \emph{incompatible} and cannot be combined. The CH approach
resolves (or evades or tames) the standard quantum paradoxes by means of the
\emph{single framework rule}, which prohibits as meaningless those arguments
which reach some conclusion by (often implicitly) combining properties
belonging to incompatible PDIs or \emph{frameworks}; the latter term can
include PDIs involving events at different times.

\xb
\outl{Frameworks $\{[\Psi],I_m-[\Psi]\}$ and $\{M_j\}$ are incompatible}
\xa

An immediate application to the present discussion arises from the fact that
the $\{M_j\}$, the possible measurement outcomes, form a PDI. But there is
another PDI with just two projectors, $[\Psi]$ and $I_m-[\Psi]$, with
$[\Psi]=\dya{\Psi}$ the weird `cat' property introduced above, and $I_m$ the
identity on the apparatus Hilbert space $\HC_m$. This PDI,
$\{[\Psi],I_m-[\Psi]\}$, is incompatible, with the $\{M_j\}$ PDI when, as we
are assuming, two or more of the $r_j$ are nonzero. Both PDIs are valid quantum
descriptions, but they cannot be combined. One cannot say, at least while being
careful in the use of language, that $[\Psi]$ corresponds to the pointer being
simultaneously in several locations, since talk of $[\Psi]$ renders talk of
pointer positions represented by the $M_j$ meaningless.%
\footnote{In the quantum logic of Birkhoff and von Neumann \cite{BrvN36}, the
  combination of $[\Psi]$ and some $M_j$ is meaningful, but false. This, alas,
  is not very helpful; see the discussion in Sec.~4.6 of \cite{Grff02c}.} %
In summary, the CH approach resolves the first measurement problem, a
Schr\"odinger cat state $\ket{\Psi}$ of the output, by employing a framework or
PDI $\{M_k\}$ as the sample space to which $\ket{\Psi}$  assigns
probabilities according to the Born rule, rather than using the PDI $\{[\Psi],
I_m-[\Psi]\}$, in which talk of the pointer having a position is
meaningless. 

\xb
\outl{2d measurement problem. Apparatus calibration.}
\xa

To justify the belief of the experimental physicist that the apparatus he has
carefully constructed to carry out some sort of measurement of a microscopic
property does what it was designed to do, we need to solve the second quantum
measurement problem. In the case of the projective measurement of $A$, this
means being able to infer from the outcome, the pointer position $M_j$, in a
particular run of the experiment, that in this run the particle actually had
the property $P_j$, corresponding to the eigenvalue $a_j$ of $A$, immediately
before the measurement took place. The theoretical analysis that justifies this
requires the use of quantum histories, and in the interests of brevity we refer
the reader to other work \cite{Grff11b,Grff15,Grff17b} for details.
Experimenters will generally want to calibrate a piece of apparatus to check
that it is working properly before using it to gather data. The simplest sort
of calibration is to send into the apparatus, on several successive runs,
particles which are known to have a specific property $P_j$, and checking that
the pointer always ends up in the corresponding position $M_j$. After doing
this for all the different $j$ values, the experimenter will feel justified in
assuming that when a particle in an unknown state arrives that causes the
pointer to end up at $M_j$, this particle just before the measurement had the
property $P_j$.

\xb
\outl{2d measurement problem if particle prepared in superposition $\ket{\psi}$}
\xa

This sounds sensible, but what if the particle was initially prepared not in an
eigenstate of $A$, but instead in some superposition, as in \eqref{eqn2}, with
several nonzero $r_j$? Granted, there is no reason, if we employ the framework
$\{M_j\}$ for analyzing the macroscopic outcome, to suppose the pointer will
end up in a weird superposition; instead it will arrive at a specific position,
which will vary from run to run. But \emph{before} the measurement took
place, did not the particle have the property $[\psi]$? That assumes a PDI
$\{[\psi], I_p-[\psi]\}$, but there is an alternative, namely the PDI
$\{P_j\}$. These two PDIs are incompatible and cannot be combined, and it is
the PDI $\{P_j\}$ which is useful in answering the question as to whether a
given measurement outcome $M_j$ revealed the prior property $P_j$. The CH
analysis justifies treating $\ket{\psi}$ in this situation as a pre-probability
which can be used to assign a probability
\begin{equation}
 p_j = \mte{\psi}{P_j}
\label{eqn5}
\end{equation}
to the property $P_j$, which will later lead (with certainty) to the measurement
outcome $M_j$. Both \eqref{eqn3} and \eqref{eqn5} are marginals
of the joint distribution
\begin{equation}
 \Pr(P_j,M_k) = \dl_{jk} p_j,
\label{eqn6}
\end{equation}
which results from a CH analysis employing histories and an extended Born rule.
Combining \eqref{eqn3} and \eqref{eqn6} yields the conditional probability
\begin{equation}
 \Pr(P_k \vbl M_j) = \dl_{jk},
\label{eqn7}
\end{equation}
which says that if the pointer ends up at $M_j$ the particle earlier had the
property $P_j$.

\xb
\section{Quantum Mechanics is Bell Noncontextual \label{sct4} }
\xa

\xb
\outl{Apparatus (Fig. 1) with $B$ vs $C$ handle can measure $A$ with $B$ or
 with $C$}
\xa

With a proper quantum-mechanical understanding of how to interpret measurements
in a way that makes sense and connects with laboratory practice, we are now in
a position to analyze the $ABC$ situation introduced in Sec.~\ref{sct2}.
Let the spectral representations for $B$ and $C$ be of the form, similar to 
\eqref{eqn1},
\begin{equation}
 B = \sum_k b_k Q_k,\quad C = \sum_l c_l R_l,
\label{eqn8}
\end{equation}
using the PDIs $\{Q_k\}$ and $\{R_l\}$, where $AB=BA$ means that $P_jQ_k=Q_kP_j$
for all $j$ and $k$, and likewise $AC=CA$ means that $P_jR_l=R_lP_j$. However,
we are assuming $BC\neq CB$, so for at least some $k$ and some $l$ it must be
the case that $Q_kR_l\neq R_lQ_k$.

\begin{figure} [h]
\hspace{2.cm}
\includegraphics[clip]{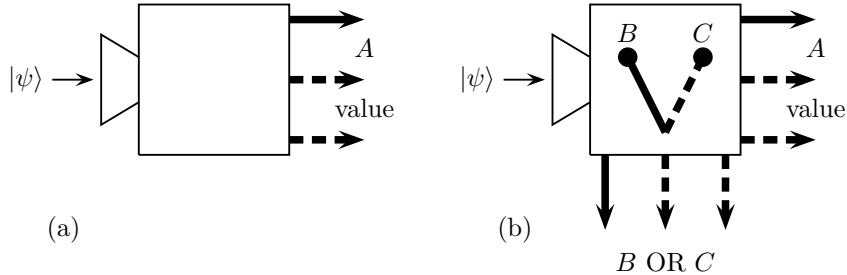}
\caption{%
  (a) Apparatus to measure $A$ for particle entering from the left, with
  outcome shown by the pointer on the right, which can be in one of several
  possible positions. (b) Apparatus which will measure $A$ together with $B$ or
  with $C$, depending on the setting of the handle, which is shown in position
  $B$, but can be moved to the $C$ position (dashed). The second pointer
  indicates the $B$ or the $C$ outcome. }
\label{fgr1}
\end{figure}

Figure~\ref{fgr1} is a schematic diagram of a measurement apparatus. In
Fig.~\ref{fgr1}(a) the particle to be measured enters from the left, the
apparatus measures $A$, and the outcome is indicated by the location of a
pointer on the right, which can be in one of several possible positions: the
solid arrow represents a particular outcome and the dashed arrows are
alternative possibilities. The modified apparatus in Fig.~\ref{fgr1}(b) has a
handle which can be set in one of two positions, $B$ or $C$. When in the $B$
position the apparatus will measure $B$ at the same time as $A$, with the value
of $B$ shown by a second pointer; while if the handle is in position $C$ it
measures $A$ together with $C$, and the second pointer indicates the measured
value of $C$. Moving the handle changes what happens inside the apparatus, and
it can be set at the very last moment, just before the particle enters on the
left side.

\xb
\outl{Calibration of measurement apparatus}
\xa

Needless to say, the careful experimenter will want to make separate sets of
calibration runs, one set with the handle in position $B$, and another
set with it in position $C$. In the first set the particles with
eigenvalues $(a_j,b_k)$ for $A$ and $B$ are repeatedly sent into the apparatus
to verify that the $A$ and the $B$ pointers arrive at the correct positions,
while the second set uses particles with eigenvalues $(a_j,
c_l)$. 

\xb
\outl{Handle at $B$, apparatus indicates $A=a_2$. Same if handle had been at
 $C$}
\xa

\xb
\outl{Refs for prior RBG ctfl pubs; RBG vs Stapp; QM is Bell noncontextual}
\xa

With the calibrations completed, consider a run in which the apparatus handle
is in position $B$ and at the end of the measurement the first pointer
indicates that the particle possessed the property $P_2$ corresponding to the
eigenvalue $a_2$, while the second pointer indicates that the particle also
had, say, the property $B=b_1$. Now suppose the handle had been in position C
rather than $B$ during this run; would the $A$ pointer nonetheless have
indicated a prior property $P_2$ corresponding to $a_2$? This is at least
plausible, because the handle could have been shifted, from $B$ to $C$, just
before the arrival of the particle, and since the particle already had the
property $P_2$ \emph{before} arriving at the apparatus, see the discussion in
Sec.~\ref{sct3}, it is hard to see how a later shift of the handle could have
altered this.
It may be worth noting that the counterfactual analysis used in the
foregoing discussion is consistent with a proposal made some years ago,
\cite{Grff99} and Ch.~19 of \cite{Grff02c}, and which has stood the test of
time; see, for example, a discussion with Stapp concerning nonlocality in
\cite{ Stpp12,Grff12b}. 

\xb
\outl{Possible retrocausal influences: not a serious concern}
\xa

But what about retrocausal influences? Could a later change of the position of
the handle have altered an earlier property of the particle? So far as I know,
there is not the slightest experimental evidence for such retrocausal
influences. Proposals for retrocausality sometimes arise out of discussions of
quantum correlations that violate Bell inequalities, along with a desire to
avoid attributing them to nonlocal influences; see, e.g., \cite{PrWh15}.
However, since such correlations have a fairly straightforward explanation in
terms of ordinary causes once one has cleared away the fog associated with
quantum measurement problems, see \cite{Grff11b,Grff19}, there seems at present
no reason to take such retrocausation proposals seriously.

\xb \outl{QM is Bell noncontextual. Claim that `Bell $\Ra$ QM 
  contextual' is an  error} \xa

Until someone points out a flaw in the argument given above, or in its
predecessors in \cite{Grff13b,Grff17b}, I will continue to maintain that
quantum mechanics is Bell \emph{noncontextual}. Authors who claim that quantum
mechanics is contextual and cite Bell's work to support their claim either have
not understood what is at issue, or have not known how to analyze quantum
measurements in a consistent  and fully quantum mechanical fashion.

\section{An Alternative Definition of `Contextual' \label{sct5}}

\xb
\subsection{Global Contextuality \label{sbct5.1}}
\xa

The term `contextual' has been used in recent years in ways different from
Bell's proposal. In some cases this is simply carelessness, but in others the
usage is distinctly different from Bell contextuality as exemplified in the
$ABC$ setup of Sec.~\ref{sct2}. The following definition of \emph{global
  contextuality} is motivated by the careful and accessible discussion in
Abramsky et al.\ \cite{AbBM17}, but it undoubtedly represents at least the
general idea behind the notion of `contextual' as used in much current work.
 
Consider a collection $\XC$ of quantum observables, all defined on the same
Hilbert space, and define a \emph{context} $\CC$ (in \cite{AbBM17} this is a
`measurement context') to be any set of \emph{compatible}---they commute with
one another---observables belonging to $\XC$. For the $ABC$ example of
Sec.~\ref{sct2}, $\XC$ is the collection $\{A,B,C\}$, while $\{A,B\}$ and
$\{A,C\}$ are the two contexts. Next assume there is a fixed initial quantum
state $\rho$ that assigns a probabilities to any projector $P$ using the Born
rule:
\begin{equation}
 \Pr(P) = \Tr(\rho P).
\label{eqn9}
\end{equation}
Using this formula will yield a joint probability distribution $\Pr_\CC$ for
the observables that make up a context $\CC$, using as a quantum
sample space all nonzero products of projectors from the PDIs
associated with the different observables in $\CC$. For example, if $A$ and
$B$, which commute with each other,
have the spectral forms given in \eqref{eqn1} and \eqref{eqn8}, their joint
probability distribution is 
\begin{equation}
 \Pr\nolimits_{\{A.B\}}(a_j,b_k) = \Tr(\rho P_j Q_k).
\label{eqn10}
\end{equation}
The marginal distribution for $B$ is obtained by
summing out $A$, 
\begin{equation}
 \Pr\nolimits_B(b_k) = \sum_j \Pr\nolimits_{\{A.B\}}(a_j,b_k) = \Tr(\rho Q_k),
\label{eqn11}
\end{equation}
and that for $A$ by summing out $B$.

This generalizes in an obvious way to any collection of commuting observables,
and allows one to define a joint probability distribution $\Pr_\CC$ for every
context $\CC$ associated with $\XC$. The collection of all these probability
distributions, one for each context, constitutes an \emph{empirical model} for
$\XC$ in the terminology of \cite{AbBM17}. The fact that these probabilities
are all generated using the same initial state $\rho$ means that they satisfy a
\emph{compatibility condition}: If two contexts $\CC$ and $\DC$ overlap, both
will assign exactly the same marginal distribution to the observables in their
intersection $\CC\cap\DC$, which itself is a context (or possibly a single
observable), even if $\CC$ and $\DC$ are incompatible in the sense that not all
observables in $\CC\cup \DC$ commute.

If all the observables in $\XC$ commute with each other, $\XC$ is itself a
context, and the probability distribution for any context $\CC$ can be obtained
as a marginal, by summing out the observables not in $\CC$, of the the joint
probability distribution $\Pr_\XC$. However, even if some observables do not
commute with others, it might still be possible to find a joint \emph{global}
distribution $\Pr_g$ for all the observables in $\XC$ such that for each
context $\CC$ the corresponding marginal of $\Pr_g$, obtained by summing out
the observables not in $\CC$, is identical to the $\Pr_\CC$ obtained using the
given $\rho$ and the Born rule \eqref{eqn9}. If such a $\Pr_g$ exists---its
entries must, of course, be nonnegative and sum to $1$---we say that the
empirical model associated with $\XC$ is \emph{globally noncontextual}, while
if such a $\Pr_g$ cannot be constructed, the empirical model is \emph{globally
  contextual}.

Thus in the case of our $ABC$ example in which $B$ and $C$ do not
commute, there might be a joint probability distribution $\Pr_g(a_j,b_k,c_l)$
whose marginals, $\Pr_g(a_j,b_k)$ and $\Pr(a_j,c_l)$ from summing out $C$ and
$B$, respectively, would coincide with the probabilities assigned to the
$\{A,B\}$ and $\{A,C\}$ contexts using a single initial state $\rho$. In this
case the empirical model, consisting of the two distributions $\Pr(a_j,b_k)$
and $\Pr(a_j,c_l)$, would be globally noncontextual, whereas if there were no
such $\Pr_g(a_j,b_k,c_l)$, the empirical model would be globally contextual.
A simple, but nontrivial, example of such a globally noncontextual model
is given in Sec.~5~(b) below. 

An example of global contextuality is provided by the well known violation of
the Bell CHSH inequality \cite{CHSH69} for two spin-half particle $a$ and $b$
with combined Hilbert space $\HC=\HC_a\ot\HC_b$. There are two noncommuting
operators $A$ (understood as $A\ot I$ on $\HC$) and $A'$ for particle $a$, and
two noncommuting operators $B$ and $B'$ for particle $b$, whereas both $A$ and
$A'$ commute with both $B$ and $B'$. The collection $\XC=\{A, A', B, B'\}$ thus
gives rise to four contexts $\{A,B\}$, $\{A,B'\}$, $\{A',B\}$ and $\{A',B'\}$.
The four probability distributions assigned to these contexts using the Born
rule for a particular initial state constitute an empirical model. For a
particular choice of operators and initial state the nonexistence of a global
joint probability distribution $\Pr_g(A,A',B,B')$ with marginals that agree
with those of the four contexts is known as Fine's theorem \cite{Fne82b}; a
discussion and some simple proofs will be found in \cite{Hllw14}.

It is important to note that global (non)contextuality as defined here applies
to an empirical model, and not directly to quantum theory. Indeed, if not all
the observables in $\XC$ commute, $\Pr_g$ cannot be a proper quantum
probability since there is no quantum sample space, no PDI, with which it can
be associated. For this reason it seems somewhat misleading to say that
``quantum mechanics is globally contextual'', even though it is true that
quantum theory makes possible empirical models that are globally contextual.

\xb
\subsection{An Example \label{sbct5.2}}
\xa

\xb
\outl{$ABC$ $\lra$ PDI's $\{P_j\}$, $\{Q_k\}$, $\{R_l\}$; probs from
empirical model ${\Pr}_g(P_j,Q_k,R_l)$ }
\xa

Let us apply the above definition to the $ABC$ case considered earlier in
Sec.~\ref{sct2}, where $\{A,B\}$ is one context and $\{A,C\}$ a different
context. Since $A$ and $B$ commute, the corresponding PDIs $\{P_j\}$ and
$\{Q_k\}$, \eqref{eqn1} and \eqref{eqn8} will have a common refinement whose
sample space consists of all the nonzero products $P_j Q_k=Q_k P_j$, and the
empirical model assigns a probability distribution 
\begin{equation}
 \Pr\nolimits_{\{A,B\}}(a_j,b_k) = \Tr(\rho P_j Q_k)
\label{eqn12}
\end{equation}
to the $\{A,B\}$ context; similarly one has
\begin{equation}
 \Pr\nolimits_{\{A,C\}}(a_j,c_l) = \Tr(\rho P_j R_l)
\label{eqn13}
\end{equation}
for the $\{A,C\}$ context. Whether this empirical model is (non)contextual
depends on the existence or nonexistence of a joint distribution
$\Pr_g(a_j,b_k,c_l)$ such that
\begin{equation}
 \Pr\nolimits_{\{A,B\}}(a_j,b_k) = \sum_l \Pr\nolimits_g(a_j,b_k,c_l),\quad
 \Pr\nolimits_{\{A,C\}}(a_j,c_l) = \sum_k \Pr\nolimits_g(a_j,b_k,c_l).
\label{eqn14}
\end{equation}

\xb
\outl{Case of $A,B,C$ as $3\tm 3$ matrices; probs from density operator $\rho$}
\xa

Consider a specific case of three operators
\begin{equation}
 A = \mat{-1 & 0 & 0\\0 & 1 & 0\\0 & 0 & 1}, \quad
 B = \mat{1 & 0 & 0\\0 & 0 & 1\\0 & 1 & 0},\quad
 C = \mat{1 & 0 & 0\\0 & 1 & 0\\0 & 0 & -1}
\label{eqn15}
\end{equation}
on a 3-dimensional Hilbert space, where the lower right $2\tm 2$ blocks of $A$,
$B$, and $C$ are, respectively, the identity, the $\sg_x$, and the $\sg_z$
Pauli matrices. Thus it is obvious that $A$ commutes with both $B$ and $C$,
whereas $BC\neq CB$. Each observable has two eigenvalues equal to $+1$ and one
equal to $-1$. If probabilities are assigned using the density operator
\begin{equation}
 \rho=\mat{ p & 0 & 0\\0 & r & 0\\ 0 & 0 & r};\quad  
p\geq 0, r\geq 0;\quad p+2r=1;
\label{eqn16}
\end{equation}
the joint probabilities of $A$ and $B$, and of $A$ and $C$, are those in the
first two boxes in the following table, where $a$, $b$, and $c$ denote
eigenvalues of $A$, $B$, and $C$. 

\xb 
\outl{Table of probabilities for $\{A,B\}$, $\{A,C\}$ and global
  probability for $\{A,B,C\}$}
\xa \vspace{.2cm}

\mbox{
\def\hsb{\hspace*{ .5 truecm plus 0.1pt minus 0.1pt}} 
\mbox{
\hsb\hsb
$\begin{array}{|c | c c|}
\multicolumn{3}{c}{\{A,B\}}\\ \hline
  b=     &1    &-1 \\ \hline
 a=1     & r   & r \\
 =-1    & p   & 0 \\ \hline
\end{array} 
$
}
\hsb

\mbox{
$\begin{array}{|c | c c|}
\multicolumn{3}{c}{\{A,C\}}\\ \hline
  c=    &1 &-1 \\ \hline
 a=1    & r   & r \\
 =-1    & p   & 0 \\ \hline
\end{array} 
$
}
\hsb

\mbox{
$\begin{array}{|c | c c c|}
\multicolumn{3}{c}{\{A,B,C\}}\\ \hline
 b=c=  &1 &-1 & b\neq c\\ \hline
 a=1     & r-s   & r-s & s\\
 =-1    & p   & 0 & 0\\ \hline
\end{array} 
$
}

\parbox{2.3cm}{
\begin{equation}
\phantom{2=2}
\label{eqn17}
\end{equation}
}

}
\vspace{.2cm}

\xb
\outl{Global $\Pr_g$ $\ra$ correct $\{A,B\}$, $\{A,C\}$ probs, 
but makes no physical sense }
\xa

The values in the third box, where $s$ can take any value between $0$ and $r$,
represent a global probability distribution $P_g$ chosen to produce the
$\{A,B\}$ and $\{A,C\}$ marginals in the first two boxes. It does not
correspond to anything in the quantum Hilbert space since, for example, it
assigns to the triple $a=1,\,b=-1,\,c=-1$ a probability $r-s$. But the
projector for $b=-1$ does not commute with the projector for $c=-1$, so there
is no subspace of the Hilbert space to which this probability can be assigned.
Also notice that $\Pr_g$ is not unique because the choice of $s$ is not
determined by the $\{A,B\}$ and $\{A,C\}$ marginals. Thus we have a situation
in which a (nonunique) global probability distribution exists, so the empirical
model consisting of the $\{A,B\}$ and $\{A,C\}$ distributions is globally
noncontextual, even though the global probability distribution $\Pr_g$ does not
correspond to anything in Hilbert space quantum physics.

\xb
\subsection{Probabilities of Measurements? \label{sbct5.3}}
\xa

\xb
\outl{Probs in empirical model refer to measurements? Of what? A Cl model is in
view}
\xa

In many papers discussing quantum contextuality, including \cite{AbBM17}, the
probabilities under discussion are not related in a direct way to microscopic
quantum properties, but are instead related to measurements in the sense of
measurement outcomes (pointer positions). But if some of the measured
observables are incompatible with others, they cannot all be measured at the
same time using a single piece of apparatus. For example, in some runs the
handle in Fig.~\ref{fgr1}(b) is at $B$, while in others it is at the $C$
position, even if the incoming particle always prepared in the same initial
state. Since the different runs are independent, they need not be
carried out in succession. One could just as well imagine two pieces of
apparatus, the first with the handle always at the $B$ location, the second
always at $C$, that simultaneously carry out measurements on two
identically-prepared particles labeled $1$ and $2$. This joint measurement can
be analyzed using an obvious extension of the approach in Sec.~\ref{sct4}, and
the two pointer readings related to a PDI, a quantum sample space, representing
properties before measurement. But now the sample space is the tensor product
of those for the individual particles $1$ and $2$, and if we choose it in such
a way that the apparatuses reveal prior properties it consists of commuting
projectors of the form
\begin{equation}
 P_{1j}Q_{1k} \ot P_{2j'} R_{2l}
\label{eqn18}
\end{equation}
in a fairly obvious notation, with subscripts $1$ and $2$ for the different
particles. This, of course, is very different from the nonexistent (from a
quantum perspective) sample space of triples that form the arguments of
${\Pr}_g()$ in \eqref{eqn14}.

One might hope to get around this difficulty by using the outcomes of
macroscopic measurements (pointer positions) to define a sample space to which
$\Pr_g$ might refer. While this provides a formal solution to the problem, it
simply conceals the difficulty that one is trying to combine the results of
different experiments carried out by different pieces of apparatus in order to
arrive at a joint probability. In general there is no unique way to do this, as
suggested by the presence of the free parameter $s$ in \eqref{eqn17}.

\xb
\outl{Cl models can $\ra$ Qm insights, but they should be labeled as Cl}
\xa

Granted, the study of probabilities not directly connected to physical
reality may nonetheless yield some useful insights as to what goes on in the
quantum world, and given the widely acknowledged conceptual difficulties of
quantum theory, new sources of insight are welcome. But to avoid adding further
confusion to the confused state of quantum foundations, it would be valuable if
discussions of this sort were to clearly distinguish ideas and interpretations
that apply directly to quantum mechanics from those in which one is, in effect,
employing classical models (in the present instance `empirical models') in
place of quantum physics.

\xb
\section{Summary and Conclusion \label{sct6}}
\xa

\xb
\outl{Analysis of Qm measurements needed to discuss Bell contextuality. QM is
  Bell noncontextual}
\xa

The careful analysis in Sec.~\ref{sct3} of what it is that quantum measurements
actually measure is needed to address the question of whether quantum mechanics
is or is not contextual in the sense originally introduced by Bell,
Sec.~\ref{sct2}: Given that $A$ was measured together with a compatible (AB=BA)
observable $B$, the $\{A,B\}$ context for $A$, would the outcome for $A$
\emph{in this particular run of the experiment} have been the same if, instead,
$A$ had been measured together with a compatible $(AC=CA)$ observable $C$, the
$\{A,C\}$ context for $A$? The question is of interest when $B$ and $C$ are
incompatible observables, $BC\neq CB$, so $A$, $B$, and $C$ cannot be measured
together in a single experiment. When an analysis of the situation is carried
out, Sec.~\ref{sct3}, using the consistent histories interpretation of quantum
mechanics, which is capable of describing the entire measurement process in
fully quantum-mechanical terms, the conclusion, Sec.~\ref{sct4}, is clear: If
the apparatus, properly constructed and calibrated, indicates that $A$ had a
particular value, say $a_1$, prior to the measurement of $A$ together with $B$,
then in this particular run the same result for $A$ would have been obtained
had $A$ been measured along with $C$, or with any other observable compatible
with $A$. Thus quantum mechanics is noncontextual, or, to be more
precise, `\emph{Bell noncontextual}', if one uses Bell's definition. It is
worth noting that were quantum theory Bell contextual it would cast grave doubt
on the results of experiments, since equipment is typically designed to measure
some specific property without concern about what other compatible quantities
might happen to be measured at the same time.

\xb
\outl{Global contextuality: $\neq$ Bell, and makes no Qml sense}
\xa

However, there are other definitions of contextuality, and that discussed in
Sec.~\ref{sct5}, to which I have given the name \emph{global contextuality},
differs from Bell contextuality in two important respects. First, rather than
referring the outcome of a particular measurement, the focus is on the
\emph{probability distribution} of outcomes of various measurements. Second,
`noncontextual' is defined in terms of the existence (and `contextual' in terms
of the nonexistence) of a joint global probability distribution for a
collection of quantum observables, \emph{whether or not they commute}. Such a
distribution, even if it exists, need not make much sense from a
quantum-mechanical perspective. The example in Sec.~5~(b) for three
observables in a 3-dimensional Hilbert space illustrates this point: there is
no quantum sample space (projective decomposition of the identity) to which
these probabilities can be assigned. If, on the other hand, one supposes that
various measurements of incompatible observables are carried out in different
runs of an experiment, of necessity using different arrangements of the
experimental apparatus, the corresponding quantum sample space, using tensor
products of quantum properties, Sec.~5~(c), is different from that
imagined in the definition of the global probability distribution. Thus while
global contextuality is a well-defined mathematical concept, it is not clear
that it represents anything of fundamental physical significance in the quantum
world. It may nonetheless be a source of useful physical insight in indicating
the extent to which quantum theory can be approximated by classical concepts,
provided the difference between classical and quantum physics is not ignored.

\xb
\outl{Bell, Global have similar notions of `contextual'. Previous discussions
  lacked Qm analysis of measurements. This is also behind nonlocality claims}
\xa

What Bell and global (non)contextuality have in common is the notion that a PDI
associated with an observable can lie in the intersection of two or more
incompatible PDIs, or `contexts' (or `frameworks' in the language of CH), and
that one should pay attention to this crucial respect in which quantum theory
differs from classical physics. What has been lacking in many previous
discussions, and which I have attempted to supply, is a description of the
simplest form of quantum measurement, a projective measurement,%
\footnote{An extension to POVMs, `generalized measurements', will be found in
  \cite{Grff17b}.} %
in order to understand what is actually measured; i.e., what the macroscopic
outcome of the measurement reveals about the microscopic state of the measured
system before the measurement took place. A large number of paradoxes in
quantum foundations arise from a failure to analyze quantum measurements in
fully quantum-mechanical terms: i.e, Hilbert space rather than, say, the
(classical) hidden variables employed by Bell and his followers. For example,
claims, based on violations of Bell inequalities, that the quantum world is
nonlocal evaporate when these classical variables are replaced by a proper
quantum analysis \cite{Grff11,Grff11b,Grff19}. Similarly, discussions of
contextuality, at least to the extent that they relate to measurements, need
to be based on a clear understanding of the quantum measurement process.

\xb
\outl{Present analysis only concerned with Qm `contextual', not psychology}
\xa

Finally, to repeat what was stated in Sec.~\ref{sct1}: the analysis presented
here is concerned only with the use of `(non)contextual' in quantum physics.
Whether or not, and if so how, these ideas carry over to the use of
(non)contextuality in psychology is something I must leave to those far more
knowledgeable about that subject than I.

\xb
\section*{Acknowledgements}
\xa 

I am grateful to Dr. Dzhafarov for the invitation to present the material found
here as a talk at the Purdue Winer Memorial Lectures in November of 2018, and
to join the other participants in contributing a written account to this
special issue of the Philosophical Transactions of the Royal Society A. Also I
am indebted to referees and to the authors of \cite{AbBM17} for a number of
critical comments.
\xb


\begin{thebibliography}{10}

\bibitem{AbBM17}
Samson Abramsky, Rui~Soares Barbosa, and Shane Mansfield.
\newblock The contextual fraction as a measure of contextuality.
\newblock {\em Phys. Rev. Lett.}, 119:050504, 2017.
\newblock arXiv:1705.07918.

\bibitem{Grff14b}
Robert~B. Griffiths.
\newblock The {C}onsistent {H}istories {A}pproach to {Q}uantum {M}echanics.
\newblock {\em Stanford Encyclopedia of Philosophy}, 2014.
\newblock http://plato.stanford.edu/entries/qm-consistent-histories/.

\bibitem{Grff11b}
Robert~B. Griffiths.
\newblock E{P}{R}, {B}ell, and quantum locality.
\newblock {\em Am. J. Phys.}, 79:954--965, 2011.
\newblock arXiv:1007.4281.

\bibitem{Grff15}
Robert~B. Griffiths.
\newblock Consistent quantum measurements.
\newblock {\em Stud. Hist. Phil. Mod. Phys.}, 52:188--197, 2015.
\newblock arXiv:1501.04813.

\bibitem{Grff17b}
Robert~B. Griffiths.
\newblock What quantum measurements measure.
\newblock {\em Phys. Rev. A}, 96:032110, 2017.
\newblock arXiv:1704.08725.

\bibitem{Grff13b}
Robert~B. Griffiths.
\newblock Hilbert space quantum mechanics is noncontextual.
\newblock {\em Stud. Hist. Phil. Mod. Phys.}, 44:174--181, 2013.
\newblock arXiv:1201.1510.

\bibitem{Bll661}
John~S. Bell.
\newblock On the problem of hidden variables in quantum mechanics.
\newblock {\em Rev. Mod. Phys.}, 38:447--452, 1966.
\newblock Reprinted in John S. Bell, \textit{Speakable and Unspeakable in
  Quantum Mechanics, 2d ed.} (Cambridge University Press, 2004), pp.~1-13.

\bibitem{Shmn09b}
Abner Shimony.
\newblock Hidden-variables models of quantum mechanics (noncontextual and
  contextual).
\newblock In Daniel Greenberger, Klaus Hentschel, and Friedel Weinert, editors,
  {\em Compendium of Quantum Physics}, pages 287--291. Springer-Verlag, Berlin,
  2009.

\bibitem{Mrmn93}
N.~David Mermin.
\newblock Hidden variables and the two theorems of {J}ohn {B}ell.
\newblock {\em Rev. Mod. Phys.}, 65:803--815, 1993.

\bibitem{BrvN36}
G.~Birkhoff and J.~von Neumann.
\newblock The logic of quantum mechanics.
\newblock {\em Ann. Math.}, 37:823--843, 1936.

\bibitem{Grff02c}
Robert~B. Griffiths.
\newblock {\em Consistent Quantum Theory}.
\newblock Cambridge University Press, Cambridge, U.K., 2002.
\newblock http://quantum.phys.cmu.edu/CQT/.

\bibitem{Grff99}
Robert~B. Griffiths.
\newblock Consistent quantum counterfactuals.
\newblock {\em Phys. Rev. A}, 60:5--9, 1999.

\bibitem{Stpp12}
Henry~P. Stapp.
\newblock Quantum locality?
\newblock {\em Found. Phys.}, 42:647--655, 2012.
\newblock arXiv:1111.5364.

\bibitem{Grff12b}
Robert~B. Griffiths.
\newblock Quantum counterfactuals and locality.
\newblock {\em Found. Phys.}, 42:674--684, 2012.
\newblock arXiv:1201.0255.

\bibitem{PrWh15}
Huw Price and Ken Wharton.
\newblock Disentangling the quantum world.
\newblock {\em Entropy}, 17:7752--7767, 2015.
\newblock arXiv:1508.01140.

\bibitem{Grff19}
Robert~B. Griffiths.
\newblock Quantum nonlocality: Myth and reality.
\newblock arXiv:1901.07050, 2019.

\bibitem{CHSH69}
John~F. Clauser, Michael~A. Horne, Abner Shimony, and Richard~A. Holt.
\newblock Proposed experiment to test local hidden-variable theories.
\newblock {\em Phys. Rev. Lett.}, 23:880--884, 1969.

\bibitem{Fne82b}
Arthur Fine.
\newblock Hidden variables, joint probability, and the {B}ell inequalities.
\newblock {\em Phys. Rev. Lett.}, 48:291--295, 1982.

\bibitem{Hllw14}
J.~J. Halliwell.
\newblock Two proofs of {F}ine's theorem.
\newblock {\em Phys. Lett. A}, 378:2945--2950, 2014.
\newblock arXiv:1403.7136.

\bibitem{Grff11}
Robert~B. Griffiths.
\newblock Quantum locality.
\newblock {\em Found. Phys.}, 41:705--733, 2011.
\newblock arXiv:0908.2914.

\end{thebibliography}
\end{document}